# Spin stiffness of chromium-based van der Waals ferromagnets

Zhixue Shu, Tai Kong

*Department of Physics, University of Arizona, Tucson, AZ 85721*

**Abstract**

Low temperature magnetization of $CrI_3$, $CrSiTe_3$ and $CrGeTe_3$ single crystals were systematically studied. Based on the temperature dependence of extrapolated spontaneous magnetization from magnetic isotherms measured at different temperatures, the spin stiffness constant ($D$) and spin excitation gap ($\Delta$) were extracted according to Bloch's law. For spin stiffness, $D$ is estimated to be $27\pm6$ meV $Å^2$, $20\pm3$ meV $Å^2$ and $38\pm7$ meV $Å^2$ for $CrI_3$, $CrSiTe_3$ and $CrGeTe_3$ respectively. Spin excitation gaps determined via Bloch's formulation have larger error bars yielding $0.59\pm0.34$ meV ($CrI_3$), $0.37\pm0.22$ meV ($CrSiTe_3$) and $0.28\pm0.19$ meV ($CrGeTe_3$). Among all three studied compounds, larger spin stiffness value leads to higher ferromagnetic transition temperature.

**Introduction**

Magnetic two-dimensional (2D) materials have attracted great attention after long-range magnetic ordering was experimentally discovered in monolayer $CrI_3$ and $CrGeTe_3$[1,2]. For 2D material research, these discoveries expand the possible tuning parameters from original charge degree of freedom to include magnetic degree of freedom. This additional tuning parameter has enabled further device capabilities and could bring in new opportunities in future microelectronics[3–5]. Recent progress in 2D magnetic materials has been primarily focusing on developing new material candidate with more functional properties[6–8], and on obtaining a better understanding of the mechanism behind such 2D long range magnetic ordering[9–11]. Among the available compounds, chromium-based van der Waals materials were most intensively studied as they were first verified to manifest 2D long range magnetic ordering. $CrI_3$, $CrSiTe_3$ and $CrGeTe_3$ belong to this family. All three compounds share very similar structures. Cr atoms are located at the center of Te or I octahedra. The edge-sharing octahedron are then arranged in a honeycomb lattice within each layer, forming a van der Waals gap between iodine in the case of $CrI_3$, and tellurium in the cases of $CrSiTe_3$ and $CrGeTe_3$. For $CrI_3$, this honeycomb layers leave an empty site in the center of the honeycomb, whereas double silicon/germanium atoms in $CrSiTe_3$ and $CrGeTe_3$ occupy these vacant centers.

Because the Wagner and Mermin theorem[12] requires magnetic anisotropy for any 2D long range magnetic ordering to exist, both magnetic exchange interaction and magnetic anisotropy play key roles in determining the magnetic property of these layered magnets. Measuring and understanding the influence of these fundamental magnetic parameters, therefore, are critical for future material design. Experimentally, bulk sample characterization provides valuable first-hand information, as an important step towards understanding and modeling magnetic behavior at low dimensions. The exchange interaction and anisotropy were previously measured by inelastic neutron scattering, ferromagnetic resonance on bulk $CrI_3$[10,11,13] and $CrSiTe_3$[14]. In this paper, we attempt to extract these parameters for $CrI_3$, $CrSiTe_3$ and $CrGeTe_3$ in the form of spin stiffness and spin excitation gap, via measuring and interpreting low temperature magnetization data based on Bloch's law[15]. Bulk magnetization measurement, as a more widely available technique, enables this approach to provide a more rapid characterization of future layered ferromagnets.

**Experimental**

CrI₃ single crystals were grown using a chemical vapor transport technique[16]. Stoichiometric starting elements (chromium powder, 99.94%, Alfa Aesar; iodine 99.8%, Alfa Aesar) were sealed in an evacuated silica tube. The silica tube was then put in an open-ended tube furnace for 3 days, where the material side was kept at 650 °C and the cooler side was kept at furnace edge, close to room temperature. CrGeTe₃ single crystals were grown using a high temperature solution growth technique. Starting elements in a molar ratio of Cr:Ge:Te = 2:6:36 (germanium, 99.999%, Alfa Aesar; tellurium, 99.999%, Alfa Aesar)[17] were packed in the Canfield Crucible Sets[18] and sealed in a silica tube under vacuum. The silica ampoule was then heated up to 800 °C over 3 hours, dwell for 10 hours and then slowly cooled to 500 °C over 60 hours, at which temperature the remaining liquid was decanted in a centrifuge. CrSiTe₃ single crystals were also grown using a solution method [19]. Starting materials with molar ratio of Cr:Si:Te = 1:2:6 (silicon, 99.999%, Alfa Aesar) were loaded in the same type of crucible-tube set-up as in CrGeTe₃. The sealed tube was then heated up to 1150 °C over 8 hours, dwell for 10 hours, and then cooled to 700 °C over 120 hours for decanting. In all cases, black, layered crystals were obtained with *c*-axis perpendicular to the layers. CrI₃ crystals were kept and handled in an argon glovebox.

Single crystal x-ray diffraction was measured using a Bruker D8 Discover diffractometer, with a microfocus and Cu Kα radiation. Layered single crystals were placed on sample pucks with facets facing up/down. Sample heights were adjusted for each measurement. Magnetization data were measured using a Quantum Design (QD) physical property measurement system (PPMS) Dynacool equipped with a vibrating sample magnetometer (vsm) (1.8 K – 300 K, 0- 90 kOe). All samples were measured with applied magnetic field parallel to their *c*-axis. Sample position drift due to vsm sample rod thermal contraction was addressed by either a periodic re-centering of the sample, or an extended waiting time (≥ 6 hours) at 1.8 K for rod length stabilization.

**Results and discussion**

All samples were first tested by single-crystal x-ray diffraction using a powder diffractometer. Fig. 1(a) shows (00L) diffractions for all compounds under study. All observed peaks are consistent with reported structure for CrI₃[16] CrSiTe₃[20] and CrGeTe₃[21]. Sharp diffraction peaks indicate good crystallinity for all samples. Ferromagnetic property was further used to verify the phase purity as well as consistency with existing literatures. In Fig. 1(b), field cooling, temperature dependent magnetization, measured under 1 kOe, are shown on the left scale. Magnetization curves are normalized to their measured values at 2 K ($M_{2K}$) for clarity. Ferromagnetic phase transition temperatures ($T_c$) were determined from the temperature derivative of magnetization, shown as dashed lines on the right scale. For $CrI_3$, the $T_c$ was determined to be ~60.1 K, consistent with reported value[16,22]. $CrGeTe_3$ and $CrSiTe_3$ have $T_c$ at 66.4 K and 34.4 K, both consistent with existing literatures[14,23].

To determine the spontaneous magnetization, or the saturation magnetic moment at each temperature, field-dependent magnetization data were collected. Take $CrI_3$ as an example. Fig. 2 shows a selection of magnetic isotherms at various temperatures between 50 kOe and 90 kOe. Low-field magnetization data show typical ferromagnetic behavior which are in good agreement with previous report[16] and thus not shown here. At higher applied magnetic fields, diamagnetic contribution due to both core diamagnetism[24] and sample holder become visible after magnetic saturation is reached. The diamagnetism is manifested by negative slopes in high field magnetization data, which happens to all samples under current study. A clear linear region in magnetic isotherms can be fitted to extrapolate the zero-field magnetization value, representing the spontaneous magnetization. In Fig. 2, dotted lines show such linear fits based on $M(T,H) = M(T, 0) - \chi H$. For the current study, we use magnetic isotherm data from 70 kOe to 90 kOe for all linear fittings, up to the temperature where the goodness of fit is above 95%.

All the extrapolated spontaneous magnetization values can then be plotted as a function of temperature, shown in Fig. 3 for $CrI_3$. Because the measured base temperature saturation moment was always within 5% of the expected 3 $\mu_B$ for $Cr^{3+}$, sample mass uncertainty that contributes to this deviation was factored by forcing all base temperature saturation magnetization to 3 $\mu_B$/Cr throughout this study.

In a simple spin wave picture under long wavelength limit, or in other words, at low enough temperatures, the magnon excitation spectrum can be approximated as $E(k) = \Delta + Dk^2$, where $\Delta$ is the $k$-independent spin wave excitation gap due to magnetic anisotropy, and $D$ is the so-called spin stiffness, which describes the $k$-dependence of the excitation, originates from exchange interaction. The magnetic anisotropy needed to induce a gap can originate from different sources, such as single ion anisotropy or anisotropic exchange interaction. In $CrI_3$, because single ion anisotropy is largely quenched, the excitation gap was proposed to come from anisotropic exchange interactions such as Dzyaloshinskii-Moriya interaction or the symmetric off-diagonal anisotropy in the Kitaev model[9,11,13]. The spin stiffness is associated with Heisenberg type, isotropic magnetic exchange interaction[9]. When the excitation gap can be ignored, the bulk temperature dependent magnetization will follow the famous Bloch's law in the form of[25]:

$$M(0,0) - M(T,0) = (\text{constant}) \times T^{3/2} \tag{1}$$

where: $M(0,0)$ is the spontaneous magnetization at zero temperature, zero field; and $M(T,0)$ is the spontaneous magnetization at a finite temperature, $T$. In a more realistic situation, considering the applied magnetic field and spin wave excitation gap, the proportional constant in front of the power law can be expressed as[15,26–28]:

$$M(0,H) - M(T,H) = g\mu_B \left(\frac{kT}{4\pi D}\right)^{\frac{3}{2}} f_{\frac{3}{2}}(\Delta'/kT) \tag{2}$$

where $f_p(y) = \sum_{n=1}^{\infty} \frac{e^{-ny}}{n^p}$ is the Bose-Einstein integral function that can account for the field-dependence of the spin excitation gap, $\Delta' = \Delta + g\mu_B H$. This method has been widely used to determine the spin stiffness in metallic ferromagnets such as iron, nickel, and their alloys[26,27,29–31], as well as other systems such as doped $LaMnO_3$[28]. In general, the estimated spin stiffness is consistent with the values obtained from more direct measurements such as inelastic neutron scattering and spin wave resonance[27]. Deviation from the general $T^{3/2}$ behavior can often be associated to higher order term in the spin wave dispersion approximation[27,32] or conduction electron's Stoner contribution[26] which will be discussed later.

In Fig. 3, the extrapolated temperature dependent, zero field spontaneous magnetization values for $CrI_3$ are fitted to both simplified gapless Eq. 1 and gapped Eq. 2. It is evident that spin excitation gap should be taken into consideration to properly interpret the low temperature magnetization behavior of $CrI_3$. Because the fitted magnetization values are zero field, spontaneous magnetization, there is no Zeeman splitting contribution to the excitation gap. Thus, the existence of an intrinsic excitation gap due to magnetic anisotropy in $CrI_3$ is compatible with the Wagner and Mermin theorem[12] that requires anisotropy to establish 2D long range magnetic ordering in $CrI_3$[1].

Similar measurements and fittings were carried out on multiple samples for all compounds under study. Fig. 4 shows obtained experimental data on $CrSiTe_3$, $CrGeTe_3$ and corresponding fitted curves. In general, fitting according to Eq. 2 works well for each dataset, whereas different samples reflect slightly different temperature dependence. Such difference may result from the variation in sample quality even within the same batch. The extracted spin stiffness constants and spin excitation gap are listed in Table 1. Error bars are calculated according to fitting results on different samples.

Since $CrI_3$ is of most intense interest recently with more experimental data, we first focus on its physical properties. From our fittings, the estimated $D$ for $CrI_3$ is 27±6 meV $Å^2$, and $\Delta$ is 0.59 ± 0.34 meV. The spin stiffness value of $CrI_3$ is close to the out-of-plane inelastic neutron scattering result[13] at $D_c \sim 23$ meV $Å^2$. For semiconducting $CrI_3$, saturation magnetization should be orientation independent. Therefore, our results reflect an averaged stiffness. Combining the reported anisotropy in magnon spectrum, the current obtained stiffness constant, however, falls short from the averaged spin stiffness[15] based on $D_{ave} = D_c^{1/3}D_{ab}^{2/3}$, where $D_{ab}$ was roughly estimated[13] to be 450 meV $Å^2$. It is unclear at this moment what contribute to this discrepancy. Future high-resolution neutron scattering measurement on in-plane magnon spectrum may help resolve the problem. The obtained spin excitation gap value shows a rather large variation among samples, despite relatively smaller difference in fitted spin stiffness values. This likely originates from high sensitivity of the Bose-Einstein integral function to the gap value. Overall, the estimated excitation gap here, $\Delta$, is qualitatively consistent with the inelastic neutron scattering and ferromagnetic resonance results of 0.3-0.4 meV[11,13]. Going back to the theoretically fitting, it was stated that the Bloch's law relies on the assumption that magnon spectrum is parabolic in reciprocal space within long wavelength regime. In the case of insulator, such as all the compounds under current study, no additional Stoner's contribution needs to be considered. Therefore, the Bloch's law is expected to work at low enough temperature. Typically, low temperature is empirically taken as below 0.2 $T_c$[28]. With a $T_c$ of ~60 K for $CrI_3$, the fitting temperature range should be smaller than 12 K, which is what has been used in Fig. 3. Furthermore, considering recent reported inelastic neutron scattering data[13], within our maximum fitting temperature in terms of energy, ~ 1 meV, the magnon spectrum exhibits a good parabolic distribution in reciprocal space. Thus, higher order term corrections are not necessary in our fitting as sometimes needed for other compounds[27,32].

CrSiTe$_3$ was also measured by inelastic neutron scattering[14]. Our results of $D = 20\pm3$ meV Å$^2$ is in good agreement with the reported value[14] at ~17 meV Å$^2$. It worth noting that the spin stiffness anisotropy is lot smaller in CrSiTe$_3$[14] than in CrI$_3$[13]. The estimated $\Delta$ from Bloch's law fitting is 0.37$\pm$0.22 meV, which is larger than reported 0.075 meV[14]. For CrGeTe$_3$, no magnon information has been reported in literature for comparison. We obtained $D = 38\pm7$ meV Å$^2$, $\Delta = 0.28\pm0.19$ for CrGeTe$_3$.

Due to the structural similarity in CrI$_3$, CrSiTe$_3$, and CrGeTe$_3$, it is physical to compare the spin stiffness, which is proportional to magnetic exchange interaction[9], to their ferromagnetic transition temperatures. In Fig. 5, $D$ is plotted against $T_c$ by black squares on the left scale. Given the uncertainty in experimental values, the $T_c$ is roughly proportional to $D$. Since spin stiffness is directly associated with the Heisenberg type isotropic interaction, these data provide a qualitative support to an intuitive picture where the stronger exchange interaction leads to higher $T_c$. It was proposed that the magnetic transition is influenced by both anisotropic and isotropic exchange[9]. However, because the large uncertainty in obtained gap values, a quantitative comparison to theory using $\Delta$ and D values here is not meaningful. Other measurements are needed to provide more accurate excitation gap values [11,13,14,33]. Although no clear correlation can be established between $\Delta$ and $T_c$, future investigation on how $\Delta$ influences the capability of stabilizing long-range magnetic ordering at 2D limit may of great interest.

**Conclusion**

In conclusion, we have synthesized single crystals of CrI$_3$, CrSiTe$_3$ and CrGeTe$_3$, three chromium-based ferromagnetic van der Waals compounds, and studied their low temperature magnetic property. Temperature dependent spontaneous magnetization values were accurately determined from magnetic isotherms measured at various temperatures. The temperature dependence of spontaneous magnetization generally follows the Bloch's law when a finite zero-field spin excitation gap is considered. Spin stiffness and excitation gap values were obtained for all three compounds. The obtained values are in general agreement with inelastic neutron scattering results for CrI$_3$ and CrSiTe$_3$, manifesting the reliability of bulk magnetization measurements in extracting spin stiffness constants. The ferromagnetic transition temperatures of CrI$_3$, CrSiTe$_3$ and CrGeTe$_3$ were observed to positively scale with their spin stiffness.

**Acknowledgments** We would like to thank X. Gui, W. Xie, V. Taufour and S. Zhang for their valuable inputs. This work is supported by University of Arizona startup funds.

**Figure Captions:**

**Figure 1**. (a) Normalized single crystal x-ray diffraction data on $CrI_3$ (black), $CrSiTe_3$ (red) and $CrGeTe_3$ (blue), showing (00L) peaks. (b) Temperature dependent magnetization of $CrI_3$ (black), $CrSiTe_3$ (red) and $CrGeTe_3$ (blue) on the left scale, and corresponding derivatives shown by dotted lines on the right scale. All values are normalized to better show ferromagnetic phase transitions in the same figure.

**Figure 2.** Field-dependent magnetization data of $CrI_3$ at a collection of representative temperatures. Dashed lines show linear fits to magnetization between 70 to 90 kOe in order to obtain spontaneous magnetization values at zero applied field.

**Figure 3.** Solid symbols show the spontaneous magnetization of $CrI_3$ as a function of temperature, extrapolated from field-dependent magnetization shown in Fig. 2. Red solid line presents model fitting according to Eq. 2 in the text, including $T^{3/2}$ and the Bose-Einstein integral. Blue dotted line shows a simple $T^{3/2}$ fitting according to Eq. 1.

**Figure 4.** Spontaneous magnetization values of (a) $CrSiTe_3$ and (b) $CrGeTe_3$ as a function of temperature. Different symbols, square, circle, and triangle represent data obtained on different samples. Solid lines with matching color are corresponding fitted curves.

**Figure 5.** Measured spin stiffness constants ($D$), spin excitation gaps ($\Delta$) as a function of ferromagnetic transition temperatures in $CrI_3$, $CrSiTe_3$ and $CrGeTe_3$.

Figure 1

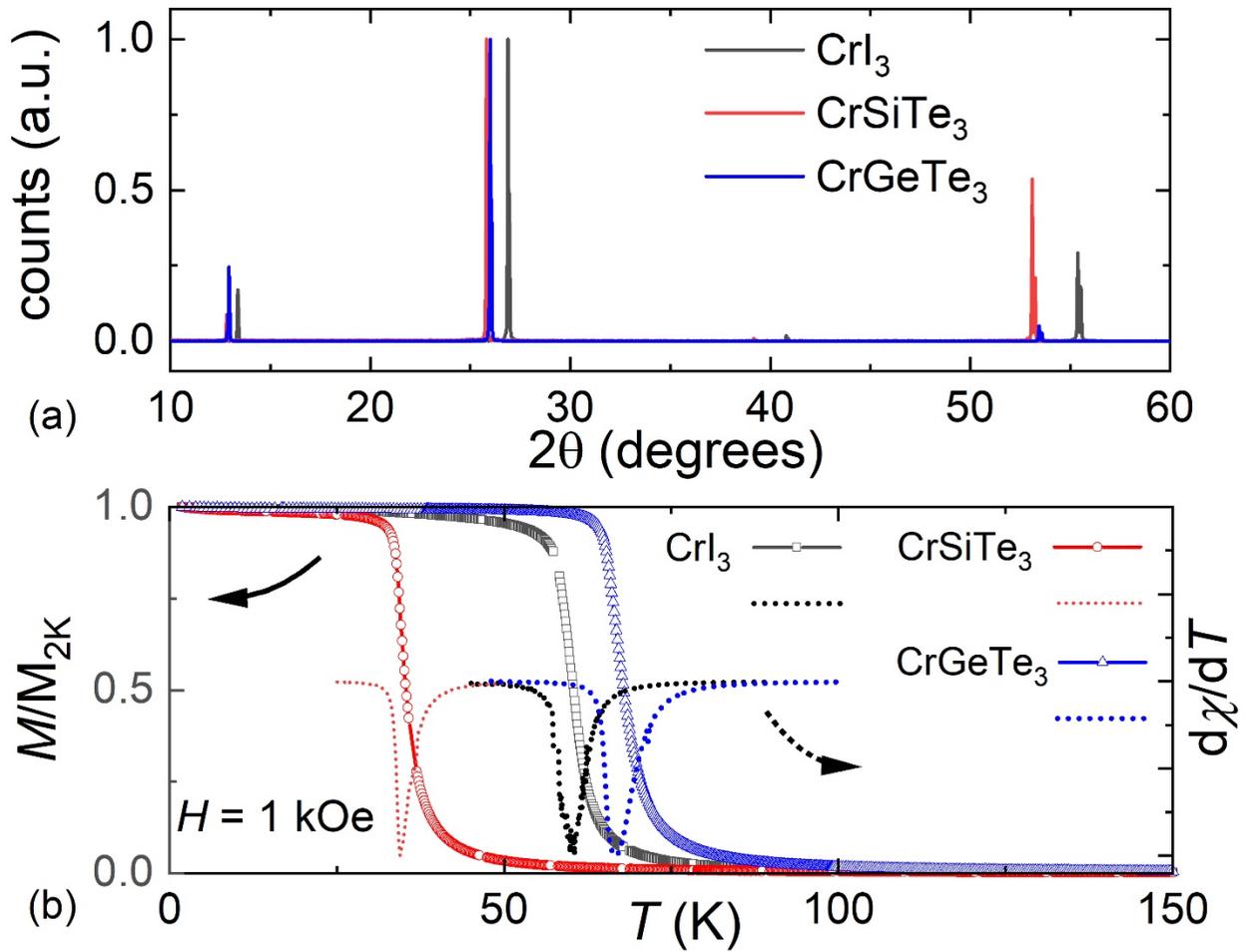

**Figure 2**

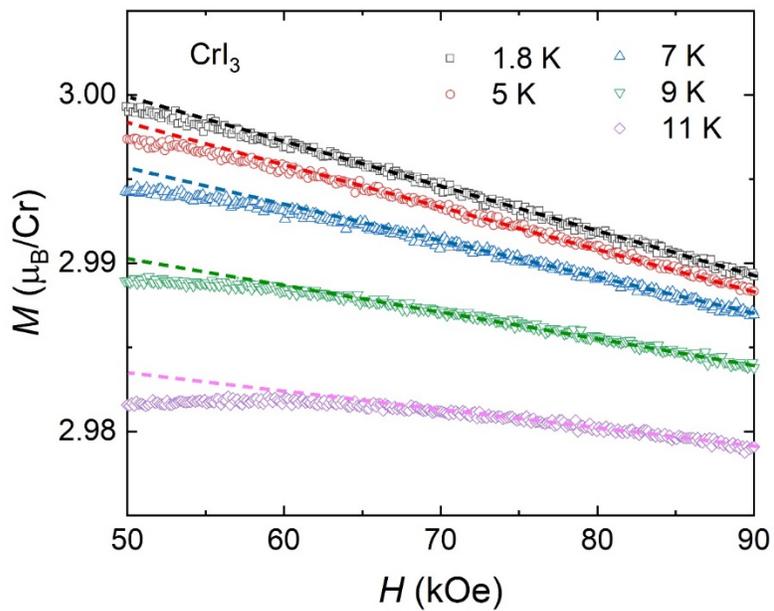

**Figure 3**

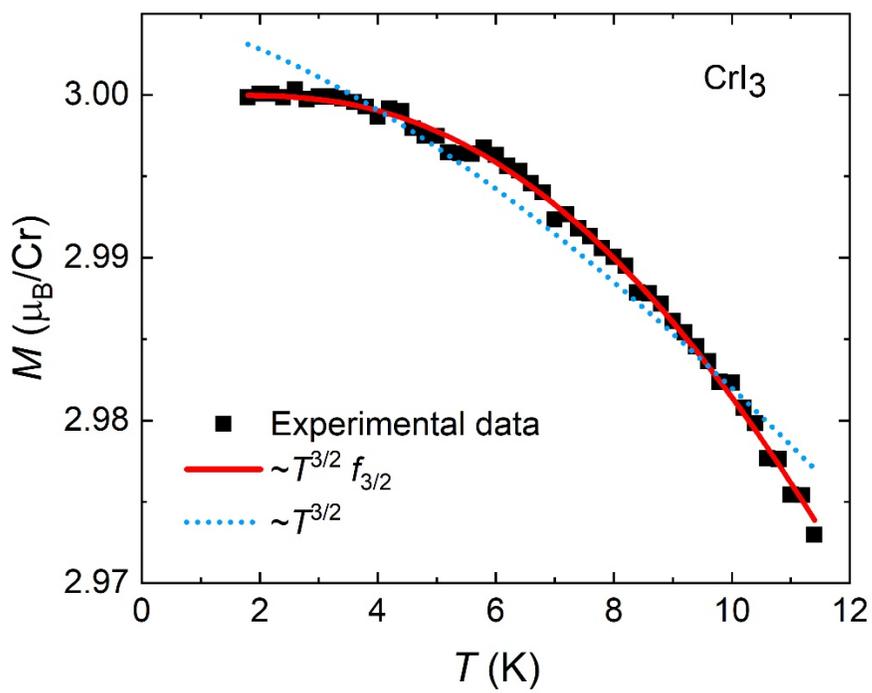

**Figure 4**

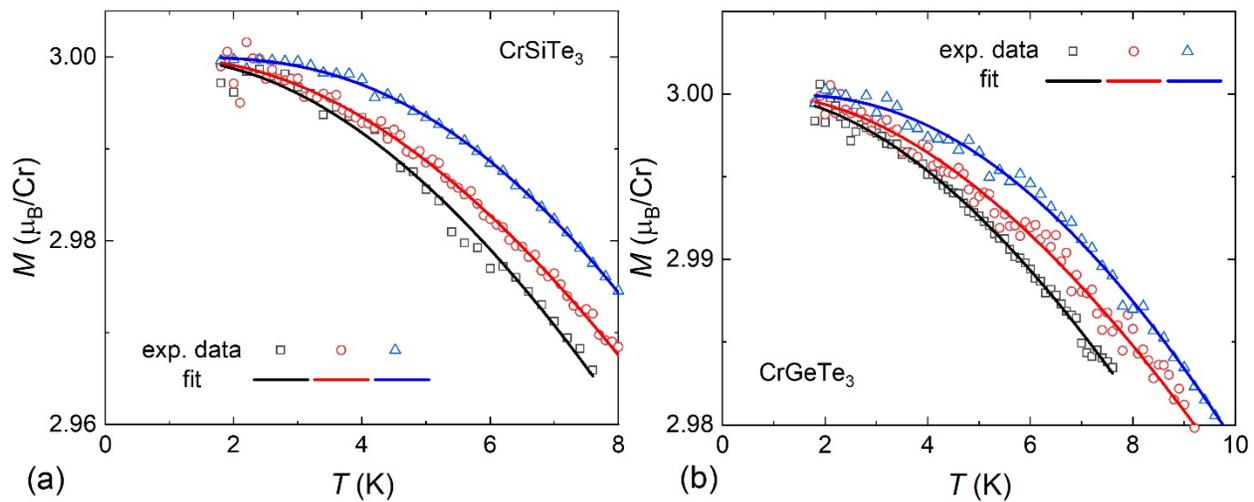

**Figure 5**

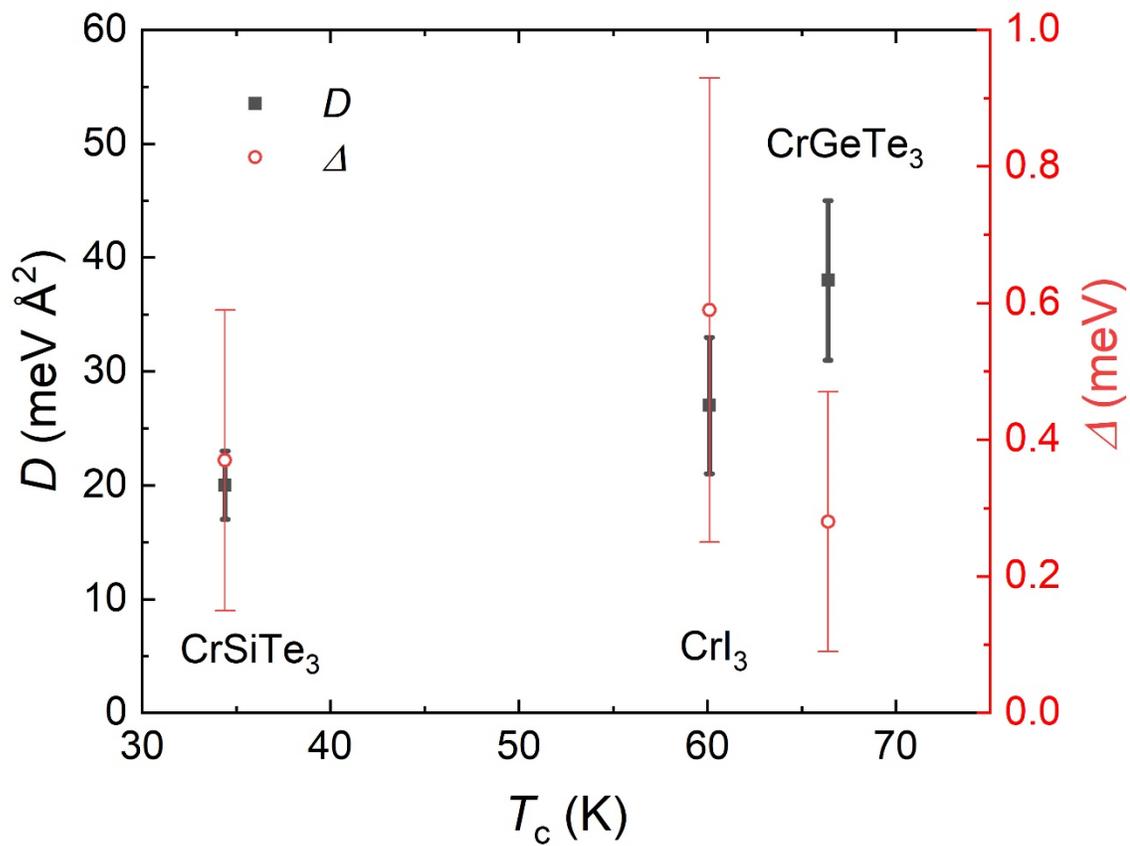

**Table 1**. Extracted spin stiffness constant $D$ and spin excitation gap $\Delta$ for $CrI_3$, $CrSiTe_3$ and $CrGeTe_3$.

| Compound | $CrI_3$ | $CrSiTe_3$ | $CrGeTe_3$ |
|---|---|---|---|
| $D$ (meV Å$^2$) | 27±6 | 20±3 | 38±7 |
| $\Delta$ (meV) | 0.59±0.34 | 0.37±0.22 | 0.28±0.19 |